\documentclass[prc,aps,reprint,superscriptaddress,showpacs]{revtex4-1}
\usepackage{amsmath}
\usepackage{graphicx}
\usepackage{url}
\usepackage{dcolumn}
\usepackage{bm}
\usepackage{units}

\begin{document}
\title{Total and partial cross sections of the $^{112}$Sn($\alpha$,$\gamma$)$^{116}$Te reaction measured via in-beam $\gamma$-ray spectroscopy}

\author{L.~Netterdon}
\email{lnetterdon@ikp.uni-koeln.de}
\affiliation{Institute for Nuclear Physics, University of Cologne, Z\"ulpicher Stra\ss e 77, D-50937 Cologne, Germany}

\author{J.~Mayer}
\affiliation{Institute for Nuclear Physics, University of Cologne, Z\"ulpicher Stra\ss e 77, D-50937 Cologne, Germany}

\author{P.~Scholz}
\affiliation{Institute for Nuclear Physics, University of Cologne, Z\"ulpicher Stra\ss e 77, D-50937 Cologne, Germany}

\author{A.~Zilges}
\affiliation{Institute for Nuclear Physics, University of Cologne, Z\"ulpicher Stra\ss e 77, D-50937 Cologne, Germany}
\begin{abstract}
\setlength{\parindent}{0pt}
\textbf{Background:} The nucleosynthesis of the neutron-deficient $p$ nuclei remains an open question in nuclear astrophysics. Beside uncertainties on the astrophysical side, the nuclear-physics input parameters entering Hauser-Feshbach calculations for the nucleosynthesis of the $p$ nuclei must be put on a firm basis.

\textbf{Purpose:} An extended database of experimental data is needed to address uncertainties of the nuclear-physics input parameters for Hauser-Feshbach calculations. Especially $\alpha$+nucleus optical model potentials at low energies are not well known. The in-beam technique with an array of high-purity germanium (HPGe) detectors was successfully applied to the measurement of absolute cross sections of an ($\alpha$,$\gamma$) reaction on a heavy nucleus at sub-Coulomb energies. 

\textbf{Method:} The total and partial cross-section values were measured by means of in-beam $\gamma$-ray spectroscopy. For this purpose, the absolute reaction yield was measured using the HPGe detector array HORUS at the FN tandem accelerator at the University of Cologne. Total and partial cross sections were measured at four different $\alpha$-particle energies from $E_\alpha\,=\unit[10.5]{MeV}$ to $E_\alpha\,=\unit[12]{MeV}$. 

\textbf{Results:} The measured total cross-section values are in excellent agreement with previous results obtained with the activation technique, which proves the validity of the applied method. With the present measurement, the discrepancy between two older data sets is removed. The experimental data was compared to Hauser-Feshbach calculations using the nuclear reaction code \textsc{TALYS}. With a modification of the semi-microscopic $\alpha$+nucleus optical model potential OMP 3, the measured cross-section values are reproduced well. Moreover, partial cross sections could be measured for the first time for an ($\alpha$,$\gamma$) reaction.

\textbf{Conclusions:} A modified version of the semi-microscopic $\alpha$+nucleus optical model potential OMP3, as well as modified proton and $\gamma$ widths, are needed in order to obtain a good agreement between experimental data and theory. It is found, that a model using a local modification of the nuclear-physics input parameters simultaneously reproduces total cross sections of the $^{112}$Sn($\alpha$,$\gamma$) and $^{112}$Sn($\alpha$,p) reactions. The measurement of partial cross sections turns out to be very important in this case in order to apply the correct $\gamma$-ray strength function in the Hauser-Feshbach calculations. The model also reproduces cross-section values of $\alpha$-induced reactions on $^{106}$Cd, as well as of ($\alpha$,n) reactions on $^{115,116}$Sn, hinting at a more global character of the obtained nuclear-physics input. 

\end{abstract}


\maketitle

\section{Introduction}
\label{sec:introduction}
The nucleosynthesis of the $p$ nuclei \cite{Woosley78,Arnould03, Rauscher13}, about 35 neutron-deficient nuclei heavier than iron bypassed by the $s$ and $r$  process \cite{Kaeppeler11,Arnould07}, is yet unclear. It is believed, that the $\gamma$ process in type II supernovae is the most dominant production chain for $p$ nuclei \cite{Rayet90,Rauscher02}. However, more processes are suggested, $e.g.$, the $rp$ process on a neutron-star surface \cite{Schatz98} or the $\nu$p process in neutrino-driven winds of type II supernovae \cite{Froehlich06}. Recently, type Ia supernovae were are also suggested as possible production sites for $p$ nuclei within the $\gamma$ process and, more efficiently, via proton-capture reactions on lighter nuclei \cite{Kusakabe11,Travaglio11}.  

The $\gamma$ process is believed to mainly take place in O/Ne layers of type II supernovae at temperatures of $ 2 \leq T \leq$~\unit[3.5]{GK}. Starting from the valley of stability, the $\gamma$ process starts with sequences of ($\gamma$,n) reactions. As the neutron separation energy decreases, ($\gamma$,p) and ($\gamma$,$\alpha$) reactions as well as $\beta$ decays will lead to deflections in the $\gamma$-process path. The reaction rates entering the $\gamma$-process reaction network are calculated within the scope of the Hauser-Feshbach model \cite{Feshbach52}. The nuclear-physics input parameters, including particle+nucleus optical model potentials (OMP), nuclear level densities and $\gamma$-ray strength functions, must be well understood. The experimental effort presented in this work aims at testing these nuclear-physics input parameters with laboratory experiments. 

Up to now, the activation technique has been the most widely used method to measure absolute reaction cross sections for charged-particle induced reactions \cite{Kiss07,Dillmann11,Sauerwein11,Halasz12,Kiss14,Glorius14,Netterdon14_1}. In addition, the 4$\pi$-summing technique is available for the investigation of certain $\alpha$- and proton-capture reactions \cite{Simon13,Quinn14}. The in-beam technique with high-purity germanium (HPGe) detectors was also successfully used for proton-capture experiments \cite{Galanopoulos03,Sauerwein12,Harissopulos13}. However, up to now no measurement using the in-beam technique with HPGe detectors has been successful in measuring the absolute cross section of an astrophysically relevant $\alpha$-induced reaction on a heavy nucleus. Recently, a dedicated setup for in-beam experiments in nuclear astrophysics became available at the Institute for Nuclear Physics in Cologne \cite{Netterdon14}. Using this setup, the $^{112}$Sn($\alpha$,$\gamma$)$^{116}$Te reaction was investigated by means of in-beam $\gamma$-ray spectroscopy for the first time. 

\section{Experiment}
\label{sec:experiment}
The $^{112}$Sn($\alpha$,$\gamma$)$^{116}$Te reaction with a $Q$ value of \unit[$\left(-962.1 \pm 28.0\right)$]{keV} \cite{qcalc} was investigated by means of the in-beam technique with HPGe detectors. Total and partial cross sections at four center-of-mass energies in the range $10.1 ~\leq~E_{\mathrm{c.m.}}~\leq~\unit[11.5]{MeV}$ were measured. Within the energy range covered in this experiment, the total cross section is dominantly sensitive to variations of the $\alpha$ width \cite{Rauscher12}. At higher energies, the total cross section becomes sensitive to the proton and $\gamma$ widths as well. The astrophysical Gamow window for this reaction is located at center-of-mass energies between \unit[$E_{\mathrm{c.m.}}~=~6.16$]{MeV} and \unit[$E_{\mathrm{c.m.}}~=~9.72$]{MeV} for a temperature of \unit[3]{GK} \cite{Rauscher10}. At this temperature, the maximum of the reaction rate integrand is located \cite{Rauscher13}. Hence, the cross section values were measured at energies slightly above the Gamow window. However, the present measurement allows stringent constraints on the nuclear-physics input parameters for the $^{112}$Sn($\alpha$,$\gamma$) reaction.

\subsection{Experimental setup}
\label{subsec:setup}

The experiment was carried out using the \unit[10]{MV} tandem ion accelerator at the Institute for Nuclear Physics at the University of Cologne, Germany. The prompt $\gamma$-rays were detected using the HPGe detector array HORUS using a setup especially designed for experiments in nuclear astrophysics \cite{Netterdon14}. The $\alpha$-particle beam with currents from 80 to \unit[240]{nA} impinged on a self-supporting $^{112}$Sn target with a thickness of \unit[$\left(364.7 \pm 14.6\right)$]{$\frac{\mu\mathrm{g}}{\mathrm{cm}^2}$}. The thickness was measured at the RUBION facility at the Ruhr-Universit\"at Bochum, Germany, by means of Rutherford backscattering spectrometry (RBS). Taking into account the enrichment of \unit[$\left(85 \pm 1\right)$]{\%} in $^{112}$Sn, this relates to an areal particle density of \unit[$\left(1.57 \pm 0.07\right) \times 10^{18}$]{$\frac{1}{\mathrm{cm}^2}$}. The energy losses of the $\alpha$-particles range from 72 to \unit[78]{keV} for the different energies. The energy losses were calculated using the \textsc{SRIM} code \cite{Srim12}. The same target was used throughout the whole experiment.

The charge deposited by the ion beam is measured at the target and at the target chamber itself. Since the beam was stopped in a thick gold backing behind the target, no charge was measured at the Faraday cup. In total, the uncertainty in the charge measurement amounts to \unit[4]{\%}. A negatively charged aperture with a voltage of \unit[$U~=~-400$]{V} prevents secondary electrons from leaving the target chamber. Moreover, the target is surrounded by a cooling trap cooled down to liquid nitrogen temperature to reduce residual gas deposits on the target. Additionally, the target chamber houses a silicon detector used for RBS measurements during the irradiation. Using this, the target thickness and stability can be monitored throughout the experiment. In the present case, no target deterioration was found within the given uncertainties.

The prompt $\gamma$ rays of the reaction products were detected using the HPGe detector array HORUS. This high-efficiency $\gamma$-ray spectrometer consists of up to 14 HPGe detectors, where six of them can be equipped with bismuth germanate (BGO) shields for an active suppression of the Compton background. In the present experiment, 13 HPGe detectors were used, five of them equipped with BGO shields. One HPGe detector and BGO shield were omitted due to geometrical reasons, in favor of mounting the RBS detector mentioned above. The distance between the detectors and target is between 9 and \unit[16]{cm} for the HPGe detectors without and with a BGO shield, respectively. The detectors are placed at five different angles with respect to the beam axis, namely at \unit[35]{$^\circ$}, \unit[45]{$^\circ$}, \unit[90]{$^\circ$}, \unit[135]{$^\circ$}, and \unit[145]{$^\circ$}. For each $\alpha$-particle energy, $\gamma$-ray spectra were additionally taken using a blank gold backing to investigate possible yield contributions from reactions occurring on the backing material. 

\begin{figure}[tb]
\centering
\includegraphics[width=\columnwidth]{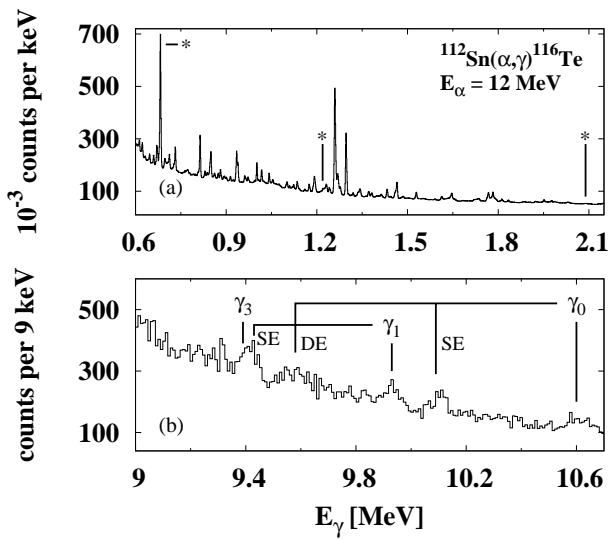}
\caption{Typical $\gamma$-ray spectrum taken during the bombardment of $^{112}$Sn with \unit[12]{MeV} $\alpha$ particles. This spectrum was obtained by summing over all HPGe detectors at an angle of \unit[90]{$^\circ$} relative to the beam axis. Ground-state transitions in $^{116}$Te are marked with an asterisk. The spectrum is dominated by beam-induced background, mainly stemming from reactions occurring on $^{56}$Fe as a target impurity. The high-energy part (b) shows the de-excitations from the compound state to the ground state ($\gamma_0$) as well as to the first ($\gamma_1$), and third ($\gamma_3$) excited state. The respective single escape (SE) and double escape (DE) peaks are marked as well, if visible.}
\label{fig:spectrum}
\end{figure}

A typical $\gamma$-ray spectrum for an $\alpha$-particle energy of \unit[$E_\alpha~=~12$]{MeV} is shown in Figs.~\ref{fig:spectrum}(a) and \ref{fig:spectrum}(b), which is the sum spectrum of five HPGe detectors placed at an angle of \unit[90]{$^\circ}$ relative to the beam axis. The spectrum is dominated by beam-induced background. However, the relevant $\gamma$-ray transitions to the ground state are clearly visible. The high-energy part of the spectrum also reveals de-excitations from the so-called entry state to the ground state and excited states in $^{116}$Te; see Fig.~\ref{fig:spectrum}(b).

\begin{figure}[tb]
\centering
\includegraphics[width=\columnwidth]{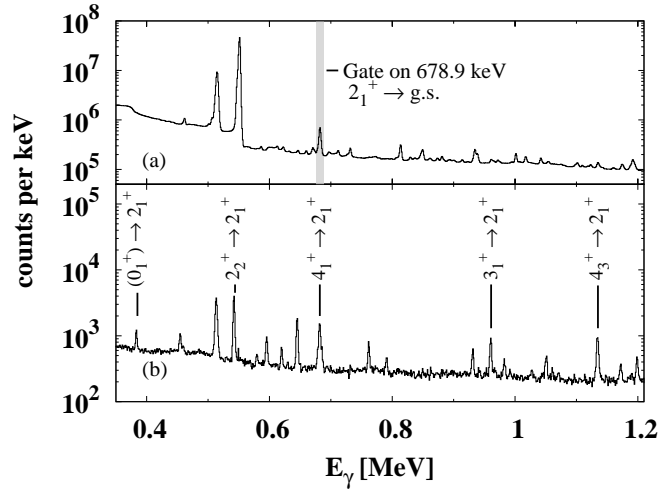}
\caption{Excerpt from a $\gamma \gamma$-coincidence spectrum of the $^{112}$Sn($\alpha$,$\gamma$) reaction with \unit[12]{MeV} $\alpha$ particles. Panel (a) shows a summed singles spectrum of the detectors positioned at \unit[90]{$^\circ$}, where no gate was applied. (b) A gate is set on the $\gamma$-ray transition from the first $J^\pi~=~2_1^+$ state to the ground state with an energy of \unit[$E_\gamma~=~678.9$]{keV}. The feeding transitions from higher-lying states in $^{116}$Te become clearly visible. Information about excitation energies, spins, and parities are adopted from Ref.~\cite{NNDC}}
\label{fig:coincidence}
\end{figure}

Due to the high granularity and detection efficiency of the setup, it is possible to measure $\gamma \gamma$ coincidences. The $\gamma \gamma$-coincidence technique is a powerful tool to suppress the beam-induced background. In the present case, where beam-induced background dominates, this is most helpful to unambiguously identify the $\gamma$-ray transitions of interest. Figures~\ref{fig:coincidence}(a) and \ref{fig:coincidence}(b) demonstrate the power of the $\gamma \gamma$-coincidence method. Figure~\ref{fig:coincidence}(a) shows a part of a $\gamma$-ray spectrum of the $^{112}$Sn($\alpha$,$\gamma$) reaction, where no gate was applied. In Fig.~\ref{fig:coincidence}(b), a coincidence spectrum is shown, after a gate on the $\gamma$-ray transition from the first excited $J^\pi~=~2_1^+$ state to the ground state was applied. The feeding transitions become clearly visible. For the relevant $\gamma$-ray transitions, no contaminants from reactions occurring on target impurities were found.

\subsection{Determination of $\alpha$-particle energy}
\label{subsec:alphaenergy}
The energy of the particle beam impinging on the target was measured by scanning the \unit[$E_p=3674.4$]{keV} resonance of the $^{27}$Al(p,$\gamma$) reaction \cite{Brenneisen95}. The proton energy was changed in small steps of 0.5 to \unit[1]{keV}. By normalizing the resonant reaction yield to the beam current, a resonance yield curve was obtained. The width of the rising edge was then used to determine the energy spread of the beam, which was found to be \unit[$\pm 3$]{keV}. Moreover, the center of this rising edge was shifted by \unit[19]{keV} with respect to the literature value. This offset can be treated as constant also for the $\alpha$ particles, since nonlinearities regarding the analyzing magnet can be excluded from earlier calibration procedures and these parameters solely depend on the geometry of the beam line, which remained unchanged during the experiment. Thus, a constant offset of \unit[19]{keV} had to be taken into account for the determination of the $\alpha$-particle energy and energy loss in the target. Details can be found in Ref.~\cite{Netterdon14}.

\subsection{Determination of full-energy peak efficiency}
The full-energy peak efficiency of the HORUS spectrometer must be precisely known up to a $\gamma$-ray energy of about \unit[10.6]{MeV} in the present case. The full-energy peak efficiency was determined using a calibrated radioactive $^{226}$Ra source for $\gamma$-ray energies up to \unit[$E_\gamma~\approx~2.5$]{MeV}. For the energy range up to \unit[$E_\gamma~\approx~3.5$]{MeV}, a $^{56}$Co source was used. The relative efficiency obtained from this measurement was scaled to the absolute full-energy peak efficiency using the \unit[$E_\gamma=846.8$]{keV} transition in $^{56}$Fe. In order to determine the full-energy peak efficiency for the highest $\gamma$-ray energies, the aforementioned \unit[$E_p=3674.4$]{keV} resonance of the $^{27}$Al(p,$\gamma$) reaction was used, which yields full-energy peak efficiencies up to a $\gamma$-ray energy of about \unit[10.5]{MeV}; for details see Ref.~\cite{Netterdon14}. 

\section{Data analysis}
\label{sec:dataanalysis}
\subsection{Reaction mechanism}
\label{subsec:compound}
\begin{figure}[tb]
\centering
\includegraphics[width=0.55\columnwidth]{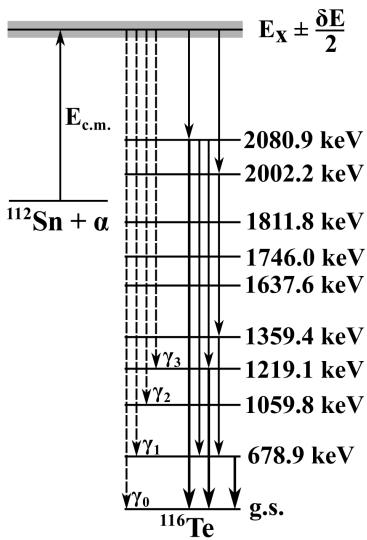}
\caption{Illustration of the excitation and decay of the compound nucleus, which is produced by bombarding $^{112}$Sn with $\alpha$ particles. The compound nucleus $^{116}$Te is formed in an excited state with energy $E_X \pm \frac{\delta E}{2}$. The entry state de-excites via $\gamma$ rays to the ground- or excited states ($\gamma_i$, depicted by dashed arrows), or by cascading $\gamma$-ray transitions to the ground state (depicted by solid arrows). The excitation energies were adopted from Ref.~\cite{NNDC}.}
\label{fig:levelscheme}
\end{figure}

The compound nuclei are produced by bombarding the target with $\alpha$ particles with an energy $E_\alpha$. In the highly excited compound nucleus, the entry state is populated with the excitation energy $E_X~=~E_{\mathrm{c.m.}}~+~Q$, where $E_{\mathrm{c.m.}}$ denotes the center-of-mass energy and $Q$ the reaction $Q$ value, which is equal to the $\alpha$-particle separation energy in the compound nucleus. Figure~\ref{fig:levelscheme} shows a partial level scheme of the compound nucleus $^{116}$Te, with illustrates the reaction mechanism as well. Within the energy uncertainty $\delta E$, a large number of unresolvable resonances are excited, upon condition that the nuclear level density is sufficiently high. This energy uncertainty is defined by the energy spread of the beam and energy straggling inside the target material and amounts to about \unit[15]{keV} for all $\alpha$-particle energies. 

From Figure~\ref{fig:levelscheme} it is obvious that the ground state can be populated either by a single transition deexciting the entry state or by cascading $\gamma$-ray transitions from higher-lying excited states. In $^{116}$Te, only three ground-state transitions from higher-lying states are known \cite{NNDC}. Using the $\gamma \gamma$-coincidence technique, see Sec.~\ref{subsec:setup}, the experimental level scheme could be verified up to the 16th excited state. No evidence for further ground-state transitions was found within this analysis.

\subsection{Determination of cross-section values}
\label{subsec:crosssections}

In order to determine the total reaction cross section, the number of produced compound nuclei $N_{\mathrm{comp}}$ must be known. This number is given by
\begin{equation}
N_{\mathrm{comp}} = \sigma \times N_{\mathrm{proj}} \times m_{\mathrm{target}}
\end{equation}
where $N_{\mathrm{proj}}$ is the number of projectiles and $m_{\mathrm{target}}$ is the areal particle density of target nuclei. $N_{\mathrm{comp}}$ is derived by measuring the absolute angular distributions of all $\gamma$ rays populating the ground state. The measured intensities $Y(E_\gamma)$ at a given angle $\theta$ are corrected for the respective number of impinging projectiles $N_p$, the full-energy peak efficiency $\varepsilon(E_\gamma)$, and the dead time of the data acquisition system $\tau$: 
\begin{equation}
W\left(\theta\right) = \frac{Y(E_\gamma)}{N_p \varepsilon(E_\gamma) \tau} \, .
\end{equation}

\begin{figure}[tb]
\centering
\includegraphics[width=\columnwidth]{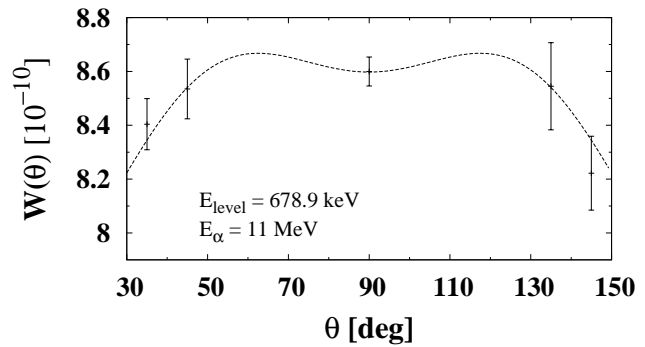}
\caption{Angular distribution for the \unit[$E_\gamma~=~678.9$]{keV} ground-state transition in $^{116}$Te. The incident $\alpha$-particle energy was \unit[11]{MeV}. The dashed line corresponds to the fit of Legendre polynomials to the experimental $W(\theta)$ values, which are calculated by normalizing the measured reaction yield to the number of incoming projectiles.}
\label{fig:angulardistribution}
\end{figure}

The angular distribution $W^i(\theta)$ of the $i$th $\gamma$-ray transition is then obtained by fitting a sum of Legendre polynomials to the five experimental values:

\begin{equation}
W^i \left(\theta\right) = A_0^i \left( 1 + \sum_{k = 2,4} \alpha_k P_k \left(\cos \theta\right) \right)
\end{equation}

with the energy-dependent coefficients $A_0$, $\alpha_2$, and $\alpha_4$. An example of an angular distribution for the $\gamma$-ray transition from the \unit[$E_X~=~678.9$]{keV} level to the ground state for an incident $\alpha$-particle energy of \unit[11]{MeV} is shown in Fig.~\ref{fig:angulardistribution}. The cross section is then calculated from the absolute coefficients of the angular distributions $A_0^i$:

\begin{equation}
\sigma = \frac{\sum_{i=1}^N A_0^i}{m_\mathrm{target}} \, ,
\end{equation}
where $N$ is the number of considered ground-state $\gamma$-ray transitions. Further details about the data-analysis procedure can be found, $e.g.$, in Ref.~\cite{Sauerwein12}. By the method of in-beam $\gamma$-ray spectroscopy and owing to the high detection efficiency of the setup, it is also possible to observe de-excitations of the compound nucleus to various excited states. For each of these $\gamma$-ray transitions, it is possible to derive the angular distributions as well in order to obtain partial cross sections. In this case, three partial cross sections could be measured, which has never been done before for an $\alpha$-induced reaction.

\section{Experimental results and discussion}
\label{sec:results}
The effective center-of-mass energies were obtained taking into account the energy loss in the target ranging from 72 to \unit[78]{keV}; see Sec.~\ref{subsec:setup}. It was determined by

\begin{equation}
E_\mathrm{c.m.} = E_{\alpha,\mathrm{c.m.}} - \frac{\Delta E}{2} \, ,
\end{equation}
were $\Delta E$ is the energy loss in the target and $E_\alpha$ the incident $\alpha$-particle energy. In the present case, the experimental cross-section uncertainties are larger than the changes of the cross-section prediction over the target thickness. Thus, this approach is valid for this experiment. The energy straggling inside the target material was approximately \unit[15]{keV} for all $\alpha$-particle energies. In order to determine the uncertainty of the $\alpha$-particle energy, the energy straggling was added to the energy spread of the $\alpha$-beam by means of Gaussian error propagation.

\subsection{Total cross sections}
\label{subsec:totalcrosssection}
\begin{figure}[tb]
\centering
\includegraphics[width=\columnwidth]{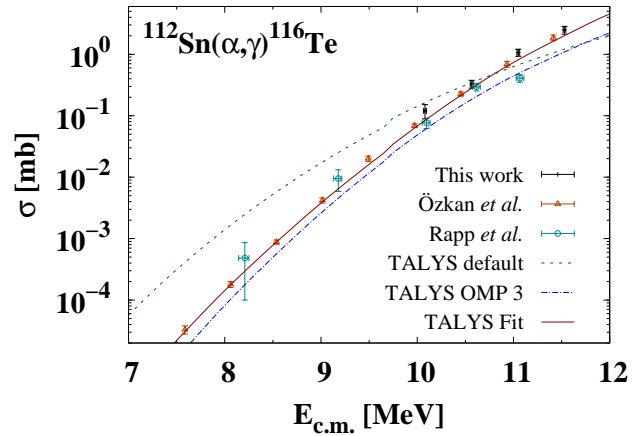}
\caption{(Color online) Experimental total cross section of the $^{112}$Sn($\alpha$,$\gamma$)$^{116}$Te reaction as a function of center-of-mass energy. Results were obtained from this work as well as from activation measurements of Ref.~\cite{Oezkan07} (\"Ozkan \textit{et al.}, triangles) and Ref.~\cite{Rapp08} (Rapp \textit{et al.}, circles). The total cross-section values are compared to statistical model calculations using the \textsc{TALYS} code. Using the default settings (`\textsc{TALYS} default'), neither the energy dependence nor the absolute values are predicted well. Using the semimicroscopic OMP 3 of Ref.~\cite{Demetriou02}, the agreement is significantly improved (`\textsc{TALYS} OMP 3'). An adjustment of the $\alpha$-OMP as well as the proton- and $\gamma$ widths leads to an excellent accordance (`\textsc{TALYS} Fit'). Details about the input parameters can be found in the text.}
\label{fig:total_xs}
\end{figure}

The experimental total cross-section values are given in Table~\ref{tab:total_xs} and shown in Fig.~\ref{fig:total_xs}. The uncertainties in this table include \unit[6]{\%} from the detection efficiency, \unit[4]{\%} from the charge collection, \unit[5]{\%} from the target thickness, and about \unit[9]{\%} to \unit[20]{\%} from statistical uncertainties.
In Figure~\ref{fig:total_xs}, the measured cross-section values from the activation measurements of Refs.~\cite{Oezkan07,Rapp08} are also shown. The presently measured cross-section values are in excellent agreement with Ref.~\cite{Oezkan07}. 

The experimental data is compared to theoretical calculations using the statistical model code \textsc{TALYS 1.6} \cite{Talys}. A comparison with a calculation using the default settings with the Watanabe $\alpha$-OMP \cite{Watanabe58} (`\textsc{TALYS} default') shows, that neither the energy dependence nor the absolute cross-section values are reproduced correctly. The agreement with a calculation applying the widely used McFadden-Satchler $\alpha$-OMP \cite{McFadden65} yields results similar to the default one, and was omitted in Fig.~\ref{fig:total_xs} in favor of better readability.
Figure~\ref{fig:total_xs} additionally shows a calculation using the semimicroscopic $\alpha$-OMP, OMP 3 of Ref.~\cite{Demetriou02} (`\textsc{TALYS} OMP 3'). Moreover, in this calculation a microscopic nuclear level density of Ref.~\cite{Goriely08} and $\gamma$-ray strength function of Ref.~\cite{Goriely04} were used, which are calculated within the Hartree-Fock-Bogoliubov (HFB) + quasiparticle random-phase-approximation (QRPA) approach. The adopted $\gamma$-ray strength function has only a minor influence on the total cross section. Using the `\textsc{TALYS} OMP 3' model, the energy dependence is well reproduced, especially at low energies. However, in order to correctly reproduce the experimental data over the whole energy region, an adjustment of the $\alpha$-OMP, as well as the proton and $\gamma$ widths, is needed. By increasing the depth of the double-folding $\alpha$-OMP, $i.e.$, the real part of the potential, by a factor of 1.16, a good description of the low-energy experimental data is obtained. For energies above \unit[$E_\mathrm{c.m.} \approx 11$]{MeV}, the proton  and $\gamma$ widths must be adjusted as well by factors of 0.2 and 1.25, respectively. With this model (denoted as `\textsc{TALYS} Fit in Fig.~\ref{fig:total_xs}), an excellent agreement with the experimental data is obtained over the whole energy region. 

\begin{table}[tb]
\caption{Experimental total cross-section values $\sigma$ of the $^{112}$Sn($\alpha$,$\gamma$) reaction for each center-of-mass energy $E_\mathrm{c.m.}$.}
\label{tab:total_xs}
\begin{ruledtabular}
\setlength{\tabcolsep}{1cm}
\begin{tabular}{rr}

\multicolumn{1}{c}{$E_\mathrm{c.m.}$ (keV)}	&	\multicolumn{1}{c}{$\sigma$ (mb)} \\

\colrule
 & \\ 
10081 $\pm$ 14 & 0.12 $\pm$ 0.03 \\
10566 $\pm$ 14 & 0.33 $\pm$ 0.05 \\
11050 $\pm$ 14 & 1.07 $\pm$ 0.14 \\
11530 $\pm$ 14 & 2.49 $\pm$ 0.36 \\

\end{tabular}
\end{ruledtabular}
\end{table}

\begin{figure}[tb]
\centering
\includegraphics[width=\columnwidth]{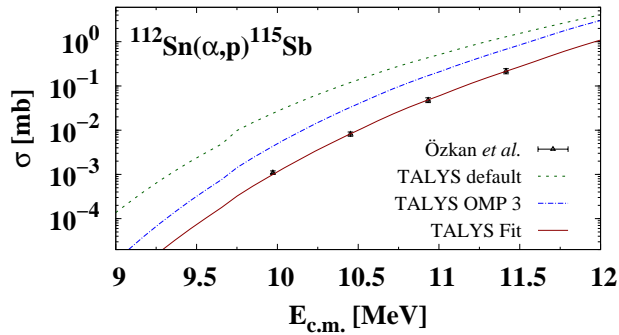}
\caption{(Color online) Experimental total cross section of the $^{112}$Sn($\alpha$,p)$^{115}$Sb reaction. The experimental data were taken from Ref.~\cite{Oezkan07}. Regarding the theoretical calculations, a pattern similar to the $^{112}$Sn($\alpha$,$\gamma$) case arises; see text for details.}
\label{fig:total_xs_a_p}
\end{figure}

The models described above were also used to calculate the total cross-section values of the $^{112}$Sn($\alpha$,p)$^{115}$Sb reaction, which were also measured using the activation technique \cite{Oezkan07}. Figure~\ref{fig:total_xs_a_p} shows a comparison of \textsc{TALYS} calculations with the experimental data. Only the adjusted model (`\textsc{TALYS} Fit') is able to reproduce the experimental data. Calculations using the other parametrizations (`\textsc{TALYS} default' and `\textsc{TALYS} OMP 3') yield a significant overestimation of the experimental cross-section values. This result strongly supports the validity of the adjusted input parameters, since both reaction channels are simultaneously well described.

\subsection{Partial cross sections}
\label{subsec:partialcrosssection}

\begin{figure}[tb]
\centering
\includegraphics[width=\columnwidth]{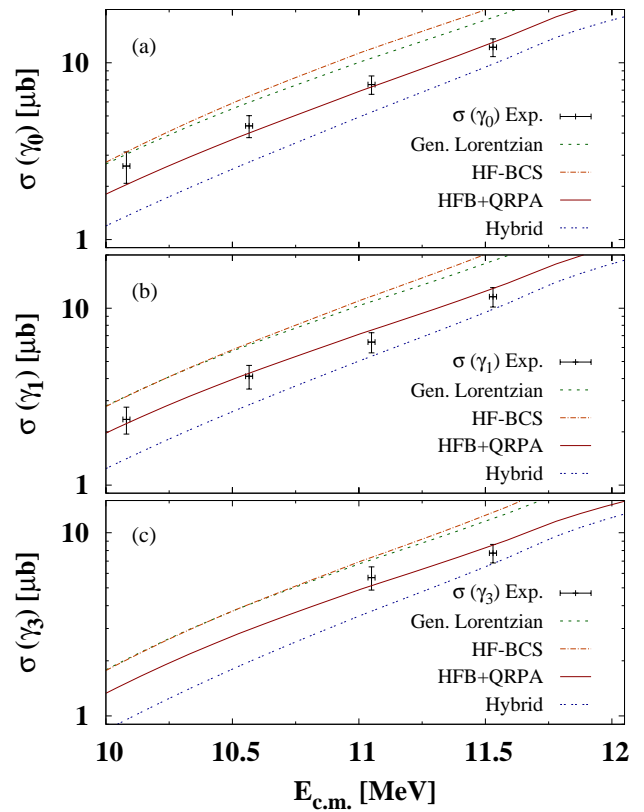}
\caption{(Color online) Experimental partial cross sections $\sigma(\gamma_i)$ for the $^{112}$Sn($\alpha$,$\gamma$) reaction as a function of center-of-mass energies. Cross-section values for the deexcitation of the entry state to the ground state, as well as first and third excited states were extracted. The experimental values are compared to \textsc{TALYS} predictions. Four $\gamma$-ray strength functions were used: generalized Lorentzian \cite{Kopecky90}, microscopic Hartree-Fock BCS \cite{Goriely02}, microscopic HFB + QRPA \cite{Goriely04}, and a microscopic hybrid model \cite{Goriely98}. Very good agreement is found using the microscopic HFB + QRPA $\gamma$-ray strength function.}
\label{fig:partial_xs}
\end{figure}

Table~\ref{tab:partial_xs} shows the experimental partial cross sections. These could be determined for the deexcitation of the entry state to the ground state as well as to the first and third excited states. For the latter, it was only possible to determine the partial cross-section values for the two highest $\alpha$-particle energies. For the lower $\alpha$-particle energies, the peak-to-background ratio was too low for a reliable determination. The partial cross sections are very valuable with respect to the $\gamma$-ray strength function. Figure~\ref{fig:partial_xs} shows a comparison of the experimental data with \textsc{TALYS} calculations. The input parameters leading to a good description of the ($\alpha,\gamma$) and ($\alpha$,p) data were used, but different models for the $\gamma$-ray strength function were adopted. In total, four different $\gamma$-ray strength functions were used for the \textsc{TALYS} calculations: generalized Lorentzian \cite{Kopecky90}, microscopic Hartree-Fock BCS \cite{Goriely02}, microscopic HFB + QRPA \cite{Goriely04}, and a microscopic hybrid model \cite{Goriely98}. The experimental partial cross sections are well reproduced by the calculation using the microscopic HFB + QRPA model. This result demonstrates the predictive power of measuring partial cross sections and, thus, the in-beam technique with HPGe detectors. By only measuring total cross-section values of the $^{112}$Sn($\alpha$,$\gamma$) reaction, no conclusion concerning the $\gamma$-ray strength function could have been drawn.

\begin{table}[tb]
\caption{Experimental partial cross sections $\sigma(\gamma_i)$ of the $^{112}$Sn($\alpha$,$\gamma$) reaction for each center-of-mass energy $E_\mathrm{c.m.}$. For the de-excitation of the entry state to the third excited state, only the cross-section values for the two highest $\alpha$-particle energies could be determined.}
\label{tab:partial_xs}
\begin{ruledtabular}
\begin{tabular}{cccc}
$E_\mathrm{c.m.}$ (keV) & $\sigma(\gamma_0)$ ($\mu$b) & $\sigma(\gamma_1)$ ($\mu$b) & $\sigma(\gamma_3)$ ($\mu$b) \\

\colrule
 & \\ 
10081 $\pm$ 14 & 2.60 $\pm$ 0.53 & 4.65 $\pm$ 0.57 & - \\
10566 $\pm$ 14 & 4.39 $\pm$ 0.63 & 5.63 $\pm$ 0.63 & - \\
11050 $\pm$ 14 & 5.52 $\pm$ 0.60 & 6.43 $\pm$ 0.84 & 5.69 $\pm$ 0.82\\
11530 $\pm$ 14 & 6.24 $\pm$ 0.82 & 7.62 $\pm$ 0.96 & 7.74 $\pm$ 0.91 \\

\end{tabular}
\end{ruledtabular}
\end{table}

\subsection{Model applicability in the Sn/Cd region}
\label{subsec:otherdata}
\begin{figure}[tb]
\centering
\includegraphics[width=\columnwidth]{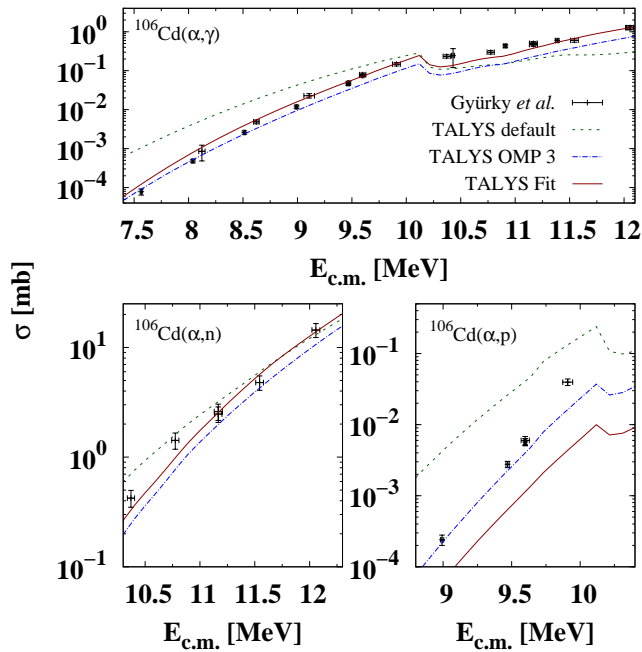}
\caption{(Color online) Total cross sections of $\alpha$-induced reactions on $^{106}$Cd. The experimental data is taken from Ref.~\cite{Gyurky06}. The models for the \textsc{TALYS} calculations are discussed in Sec.~\ref{subsec:totalcrosssection}. The model fitted to the $\alpha$-induced reactions on $^{112}$Sn (`\textsc{TALYS} Fit') yield a reasonable agreement of the $^{106}$Cd($\alpha$,$\gamma$) and $^{106}$Cd($\alpha$,n) cross sections, whereas the $^{106}$Cd($\alpha$,p) cross-section values are significantly underestimated.}
\label{fig:total_xs_cd}
\end{figure}

Motivated by the success of predicting total cross sections of the $^{112}$Sn($\alpha$,$\gamma$) and $^{112}$Sn($\alpha$,p) reactions, as well as partial cross sections of the $^{112}$Sn($\alpha$,$\gamma$) reaction, the model as described above was used to calculate cross sections of $\alpha$-induced reactions on $^{106}$Cd. Cross sections for the $^{106}$Cd($\alpha$,$\gamma$)$^{110}$Sn, $^{106}$Cd($\alpha$,n)$^{109}$Sn, and $^{106}$Cd($\alpha$,p)$^{109}$In reactions have been measured using the activation method \cite{Gyurky06}. Figure~\ref{fig:total_xs_cd} shows a comparison of the experimental data of Ref.~\cite{Gyurky06} with \textsc{TALYS} predictions. As input parameters, the same models as discussed in Sec.~\ref{subsec:totalcrosssection} were used. A similar pattern arises as for the $^{112}$Sn + $\alpha$ case. The experimental data of the $^{106}$Cd($\alpha$,$\gamma$) and $^{106}$Cd($\alpha$,n) reactions are reasonably well reproduced using the model fitted to the $^{112}$Sn + $\alpha$ reactions. However, the ($\alpha$,p) channel is significantly underestimated. The $^{106}$Cd($\alpha$,p) cross section shows a complicated sensitivity to the $\alpha$, $\gamma$, and proton widths \cite{Rauscher12}. Since the other reaction channels are rather insensitive to changes in the proton width, a deficiency in this nuclear-physics input is most probably the reason for the disagreement between experiment and theory in this case. Nevertheless, one can conclude that the modified $\alpha$-OMP is also valid for $\alpha$-induced reactions on $^{106}$Cd.

\begin{figure}[tb]
\centering
\includegraphics[width=\columnwidth]{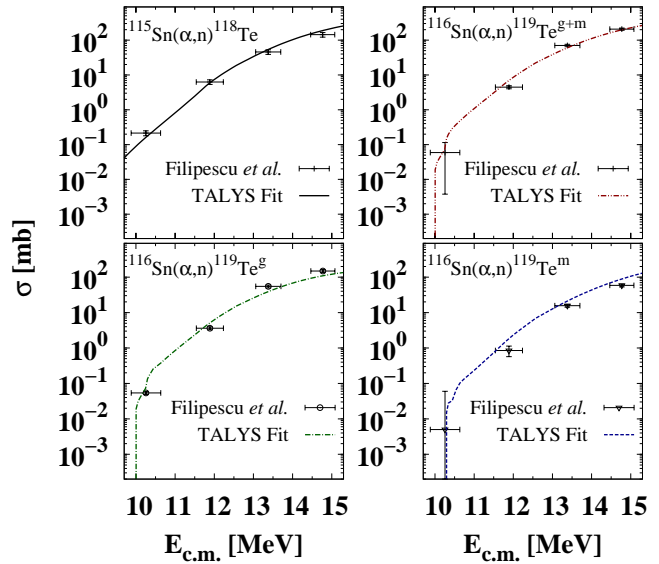}
\caption{(Color online) Cross sections of ($\alpha$,n) reactions on the isotopes $^{115,116}$Sn. Experimental data is taken from Ref.~\cite{Filipescu11}. The model (`\textsc{TALYS} Fit'), see Sec.~\ref{subsec:totalcrosssection}, is able to correctly reproduce the total cross-section values of the $^{115}$Sn($\alpha$,n)$^{118}$Te reaction. For the $^{116}$Sn($\alpha$,n)$^{119}$Te case, minor differences arise at higher energies in describing the population of the ground state and isomeric state. However, the total cross-section values are in excellent agreement with the experimental data.}
\label{fig:total_xs_sn}
\end{figure}

The obtained model was additionally tested on other $\alpha$-induced reactions along the Sn isotopic chain, namely $^{115,116}$Sn($\alpha$,n)$^{118,119}$Te. Experimental data are available from an activation measurement from Ref.~\cite{Filipescu11}, which are shown in Fig.~\ref{fig:total_xs_sn}, compared to \textsc{TALYS} calculations. For the $^{115}$Sn($\alpha$,n)$^{118}$Te reaction, an excellent agreement between experimental data and theoretical predictions is found. In the case of the $^{116}$Sn($\alpha$,n)$^{119}$Te reaction, cross-section values for the population of the ground state and the isomeric state are available. For the higher energies, the ground-state cross section is slightly underestimated, whereas the population of the isomeric state is overestimated. However, the total cross section is reproduced correctly. Thus, the model that was fitted locally to $\alpha$-induced reactions on $^{112}$Sn is also capable of describing $(\alpha$,n) reactions on the Sn isotopes $^{115,116}$Sn.

\section{Summary and conclusion}
Total and partial cross sections of the $^{112}$Sn($\alpha$,$\gamma$)$^{116}$Te reaction have been measured by means of the in-beam technique with HPGe detectors at four center-of-mass energies between \unit[$E_{\mathrm{c.m.}}~=~10.05$]{MeV} and \unit[$E_{\mathrm{c.m.}}~=~11.53$]{MeV}. The high-efficiency HPGe detector array HORUS was used for this purpose. For the first time, it was possible to investigate an ($\alpha$,$\gamma$) reaction on a $p$ nucleus using this method. Besides the total cross-section values, partial cross sections for the deexcitation to the ground state as well as to the first and third excited state were measured. An adjustment of the semi-microscopic $\alpha$-OMP of Ref.~\cite{Demetriou02} as well as of the proton and $\gamma$ widths is needed to correctly reproduce the experimental data with \textsc{TALYS} calculations. The partial cross sections are crucial in this case to apply the correct $\gamma$-ray strength function in the statistical model calculation, which is calculated microscopically within the HFB + QRPA approach \cite{Goriely04}. The presently used method, which allows one to measure cross sections of ($\alpha$,$\gamma$) reactions with a stable reaction product, is able to widely extend the experimental possibilities towards a more complete experimental data base for $\gamma$-process nucleosynthesis. A reaction worth being measured within this scope is, $e.g.$, the $^{108}$Cd($\alpha$,$\gamma$)$^{112}$Sn reaction. A systematic comparison with other experimental data in the Sn / Cd region shows, that the presently adjusted model is capable of describing $\alpha$-induced reactions on $^{106}$Cd, as well as ($\alpha$,n) on the Sn isotopes $^{115,116}$Sn. Hence, this model is, to some extent, of global character, which is worth being tested on astrophysically relevant $\alpha$-induced reactions in other mass regions as well.

\begin{acknowledgments}
The authors thank A. Dewald and the accelerator staff at the Institute for Nuclear Physics at the University of Cologne for providing excellent beams. Additionally, the authors gratefully acknowledge K.~O.~Zell for the target preparation as well as H.~W. Becker of the Ruhr-Universit\"at Bochum for the assistance on RBS measurements. This project was partially supported by the Deutsche Forschungsgemeinschaft under contract DFG (ZI 410/5-1) and the ULDETIS project within the UoC Excellence Initiative institutional strategy.
\end{acknowledgments}

\end{document}